\documentclass[]{spie}  
\usepackage[]{graphicx}
\usepackage[normalem]{ulem} 
\useunder{\uline}{\ul}{}

\title{A new pulse shape description for {\LARGE $\alpha$} particle pulses in a highly-sensitive sub-Kelvin bolometer} 

\author{S. L. Stever\supit{a,b}, F. Couchot\supit{b}, N. Coron\supit{a}, R. M. J. Janssen\supit{a}, B. Maffei\supit{a}
\skiplinehalf
\supit{a}Institut d'Astrophysique Spatiale, \textit{INSU/CNRS}, Bt. 121, Univ. Paris-Sud, Orsay, 91405, FR \\
\supit{b}Laboratoire de l'Acc\'el\'erateur Lin\'eaire, \textit{IN2P3/CNRS}, Bt. 200, Univ. Paris-Sud, Orsay, 91405, FR
}

\authorinfo{Further author information: (Send correspondence to S. L. Stever)\\E-mail: sstever@ias.u-psud.fr, Telephone: +33 (0)1 69 85 86 96}

\begin{document} 
  \maketitle 

\begin{abstract}
The next generation of cosmology space missions will be sensitive to parasitic signals arising from cosmic rays. Using a composite bolometer, we have investigated pulses produced by $\alpha$ particles in order to understand the movement of energy produced by ionising radiation. Using a series of measurements at 100 mK, we have compared the typical fitting algorithm (a mathematical model) with a second method of pulse interpretation by convolving the detector's thermal response function with a starting profile of thermalised athermal phonons, taking into account the effects of heat propagation. Using this new fitting method, we have eliminated the need for a non-physical quadratic nonlinearity factor produced using more common methods, and we find a pulse form in good agreement with known aspects of thermal physics. This work is carried forward in the effort to produce a physical model for energy deposition in this detector. The modelling is motivated by the reproduction of statistical features in the experimental dataset, and the new interpretation of $\alpha$ pulse shapes represents an improvement in the current understanding of the energy propagation mechanisms in this detector.
\end{abstract}



\section{INTRODUCTION}
\label{sec:intro}  

Following the launch of European Space Agency's Planck mission\cite{lamarre2010planck}, the susceptibility of highly-sensitive, sub-Kelvin detectors to parasitic signals from cosmic radiation has become a topic of great interest. Planck was launched to orbit at the second Earth-Sun Lagrange point (L2), and its High Frequency Instrument (HFI) contained 54 bolometers with an NEP to the order of 10$^{-17}$ W/$\sqrt{Hz}$ at 100 mK. For its time, this was the lowest detector temperature achieved in space, and it resulted unprecedented sensitivity for the mission. However, increased sensitivity to cosmic signal also meant increased sensitivity to unwanted systematic effects. Planck's time-ordered data, across all bolometers, had an unexpected number of thermal spikes which arose from impacts by cosmic rays. The empirical study and extraction of this signal was a matter of great effort for the cosmology community, and was eventually successful owing to a combination of ground-based experimental work and in-situ analysis on Planck bolometer data\cite{catalano2014characterization}$^{,}$\cite{catalano2014impact} . \\

As the scientific requirements increase in the next generation of missions, the `cosmic ray problem' will only continue to grow as devices become more sensitive and operate at even lower temperatures. Furthermore, the next generation of missions seek to measure even smaller cosmic signals, and will require a detailed understanding of the subsystem vulnerability to cosmic radiation, as well as the effects this could have on the data itself. Planck's pulse removal was successful in characterising the physical origin of the cosmic ray pulses in the data. However, future missions must take into account -- and attempt to minimise --  the potential impact of this important systematic effect, particularly during the early phases of development. \\

To this end, we have been developing a physical model for the deposition of energy in similar high-sensitivity detectors, with the motivation of understanding the mechanisms of energy propagation and eventual reproduction of the pulses. This is a complex issue, but one which is necessary to understand and predict the extent to which future missions can be vulnerable to cosmic rays. We have based this model on several rounds of experimental data using $\alpha$ radiation on a composite semiconductor bolometer, which have reproduced certain statistical features. The reproduction of these statistical features motivates the development of the model - it is a necessary tool which allows us to understand the minutiae of the behaviour of this detector. The purpose of this paper is to present an updated view of our interpretation of the pulse shape, based on a new set of data taken at 100 mK, which we compare with analysis using more conventional methods. We will compare the typical mathematical pulse shape with a new generic fitting function motivated by relevant aspects of thermal physics.\\

In this manuscript, we will outline the experimental details in Sec.~\ref{sec:expo}. We will then describe the methods of data analysis in Sec.~\ref{sec:methods}, starting with the preprocessing of the data and the verification of $\alpha$ pulses. We will describe a typical pulse analysis function used in the literature (and its limitations) in Sec.~\ref{sec:oldfits}, where we will also show some of its results in this dataset as a basis for comparison with the new model, which is described in detail in Sec.~\ref{sec:fitalgo}. The results of this new algorithm will be shown in Sec.~\ref{sec:analysis} and discussed in Sec.~\ref{sec:discussion}.\\

\section{Experimental Simulation of Cosmic Rays} 
\label{sec:expo}

\begin{figure}[htbp]
\centering
\includegraphics[width=0.7\linewidth, keepaspectratio]{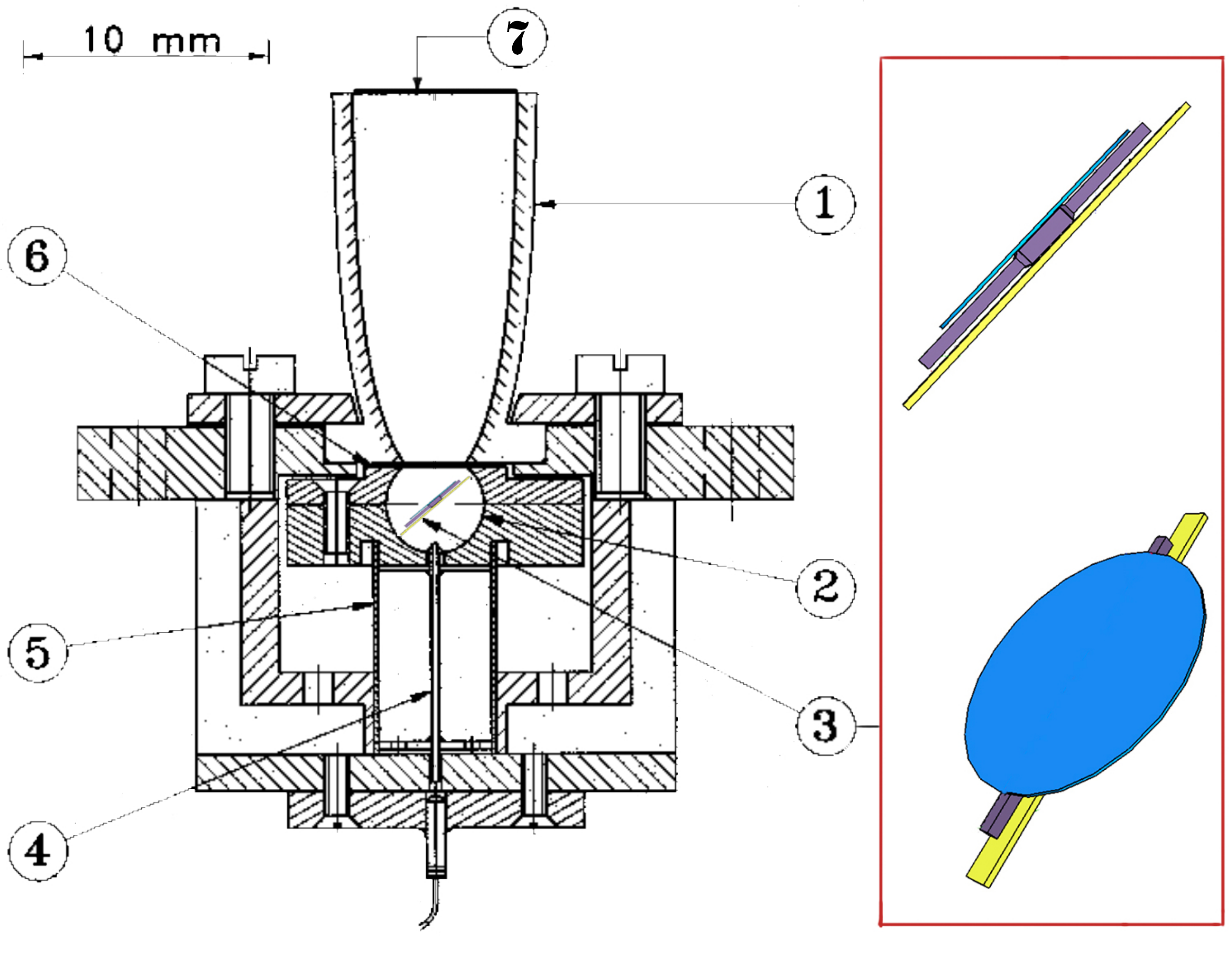}
\caption{Diagram of the bolometer used in this study, including its enclosure. \textit{Left}: The bolometer housing (figure adapted from Benoit et al.\cite{benoit2000calibration}): (1) conical feedhorn; (2) integrating sphere ($\phi$ = 4 mm); (3) bolometer (detail at right); (4) optical fibre for control of response stability, disconnected for this experiment; (5) low-pass thermal filter, also disconnected; (6) 50$\mu$m gap in thermal filter; (7) the radioactive $^{241}$Am source. Below the source is a diaphragm of $\phi$ = 2.5 mm mm, below which a second diaphragm of $\phi$ = 2 mm is separated by a distance of 12.5 mm. \textit{Right}: Bolometer details. The diamond absorber (blue) is 40 $\mu$m thick with a thin back-layer of bismuth (to absorb IR), and is affixed to a NTD germanium sensor (purple) with thin legs, and affixed to a slab of sapphire (yellow). Each leg of the sapphire and Ge are coupled to the thermal bath with wound copper wire.}
\label{DBsketch2}
\end{figure}

To produce spacelike conditions in the laboratory, we use a composite semicondcutor bolometer in a dilution-cooled cryostat, and we produce pulses using internal $\alpha$ radiation. The temperature of the cryogenic test plate is set to 100 mK using a resistor heater and a PID, and the bias voltage is set to V$_{bias}$ = 0.25 V to measure pulses in the linear regime of the VI curve. We simulate cosmic rays using $^{241}$Am $\alpha$ particles with an energy of .8 pJ, and measure them on an NTD germanium semiconductor bolometer from the ground-based DIABOLO experiment~\cite{benoit2000calibration}, referred to hereafter as `Bolo 184'. The bolometer is comprised of a diamond absorber with a thin back-layer of bismuth, a long and thin NTD germanium sensor, with a long and thin sapphire slab below. The bolometer is coupled to the thermal bath at both ends of the sensor and the sapphire, and the layers are affixed to each other using a thin layer of Devcon epoxy. The layout of the detector and its housing are shown in Fig.~\ref{DBsketch2}.\\

A total of 620 $\alpha$ particle events were recorded, although it is important to note that at 100 mK and V$_{bias}$ = 0.25 V, the detector is sensitive enough to also record cosmic ray events. To account for this, the oscilloscope trigger voltage was set to be just below the typical $\alpha$ particle amplitude. Remaining cosmic rays in this data set can be removed manually based on interpretation of the pulse attributes.\\

It is noted that in the overall goal of simulating space-like conditions, it will be necessary to use a much wider range of particles and energies than 5.4 MeV $\alpha$ particles alone. The vast majority of cosmic rays at L2 are expected to be relativistic protons, and a new test system is presently being developed for the purpose of measuring the behaviour of detectors in the beam line of a particle accelerator~\cite{Janssen2018}, although using an internal conversion-electron source is another option which we are investigating. Whilst this study is presently experimentally limited, we note that studies using $\alpha$ particles were commonly employed in Planck\cite{catalano2014characterization}$^{,}$~\cite{catalano2014impact} and provide several advantages; the higher ionising power of $\alpha$ particles result in all of the energy being embedded in the uppermost area of the first surface they encounter (in the case of Bolo 184, the upper 14 $\mu$m of the diamond absorber) which allows for the definitive understanding of where that thermal energy is placed and how it moves in the bolometer.\\

With modelling in mind as a motivation, the simplified case of localised energy deposition is simpler to reproduce, which can later be expanded on in future modelling for minimally-ionising particles, which will traverse multiple detector regions in the line-of-sight of their movement. \\

Low-amplitude cosmic ray events were also measured at 100 mK, and while they are outside of the scope of this manuscript, they will be used in parallel analysis to confirm the current understanding of the pulse shape in this detector. The data used in this manuscript is the last data set taken using the older system, which has since been decommissioned. \\


\section{Methods of Analysis} 
\label{sec:methods}

In this section, we outline the process of data analysis. We first analyse the amplitude and energy spectra of 5.4 MeV $\alpha$ particle impacts in this detector, which allows us to verify the presence of $\alpha$ particles. We then describe the data processing pipeline, which includes preprocessing procedures, a commonly-used mathematical model for describing the pulse shape (and some of its results), along with a new physically-motivated fitting algorithm for interpreting the movement of energy in this detector.

\subsection{Data Preprocessing} 
In order to finely resolve the important physical characteristics of the glitches, a sampling rate of 1 sample every 4 $\mu$s was used in experimental data acquisition. Due to the lowered bias voltage, higher sensitivity, and high sampling rate, the pulses had significant noise, with an average noise RMS of 0.067 mK. In an effort to reduce noise and computational time, the pulses were processed before fitting with uneven downsampling. The regions of the glitch with the most information (e.g. the initial athermal rise and decay points) had higher sampling than regions with less information (e.g. initial rise and decay times had 1:1 sampling whilst thermal decay areas had 1:20 or 1:50 sampling). For calculating the noise levels on the pulses, we take the RMS value of the first 2500 (of 25002) points  -- buffered data regions before the pulse trigger -- which is divided by the square root of the sampling factor to produce a flat error with respect to the sampling.\\

\begin{figure}[htbp]
\centering
\includegraphics[width=0.95\linewidth, keepaspectratio]{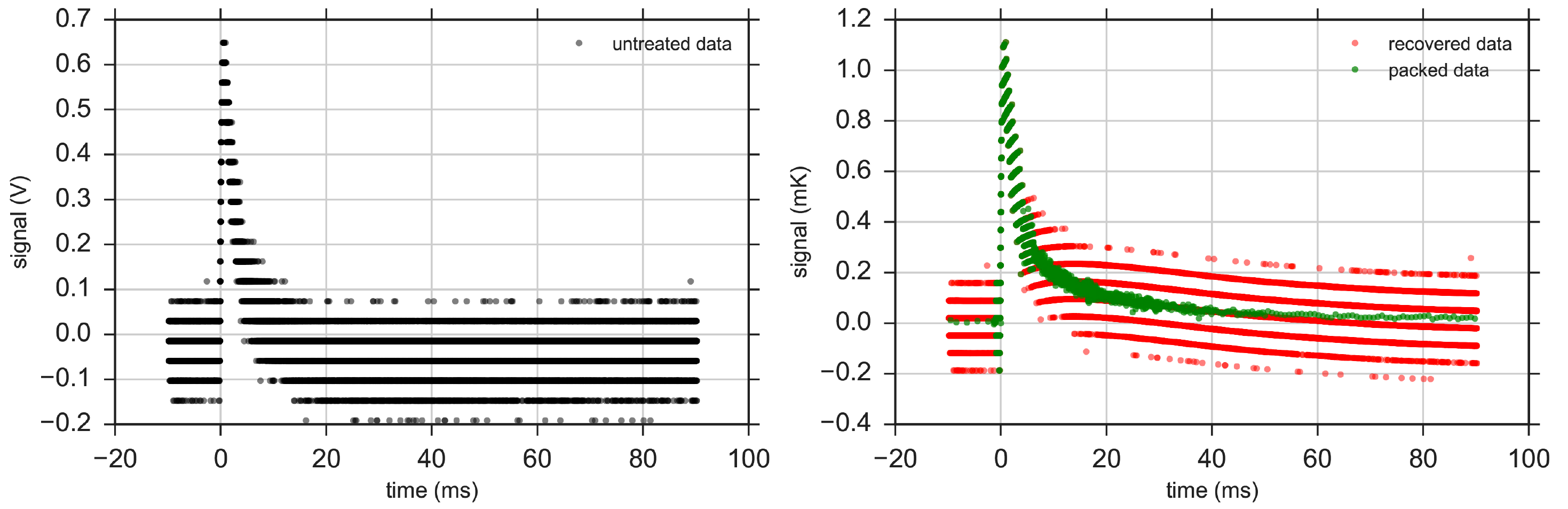}
\caption{One sample pulse, showing the data processing stages. \textit{Black}: Untreated AC-coupled pulse signal, in V; \textit{Red}: Recovered pulse signal, in mK; \textit{Green}: Unevenly sampled (packed) pulse, in mK.}
\label{dataproc}
\end{figure}

The pulses were taken on a 10 M$\Omega$ AC-coupled oscilloscope, which changes the pulse shape (specifically, it adds a non-physical negative component to the signal). This introduces a significant systematic effect into the data - however, since this data is from the final measurement run on this test system, it must be reversed via preprocessing rather than by re-measuring. To remove this effect, we use an algorithm which integrates the signal to regenerate an equivalent DC-coupled input -- this has the drawback of introducing a small artificial slope, which we account for in the next section. We apply an additional reversal filter to remove the effect of the .1 Hz high-pass filter on the pre-amplifier, and a third algorithm which converts the pulses (in volts) to units of resistance, and then into temperature based on the R(T) characteristics of the detector (which are based on measurements). The final output of the preprocessing is T(t) curves in mK with uneven temporal sampling. These are the curves used in the remainder of this analysis. A sample pulse demonstrating each stage of the pulse processing is shown in Fig~\ref{dataproc}.

\subsection{Amplitude and Energy Spectra}

\begin{figure}[htbp]
\centering
\includegraphics[width=0.9\linewidth, keepaspectratio]{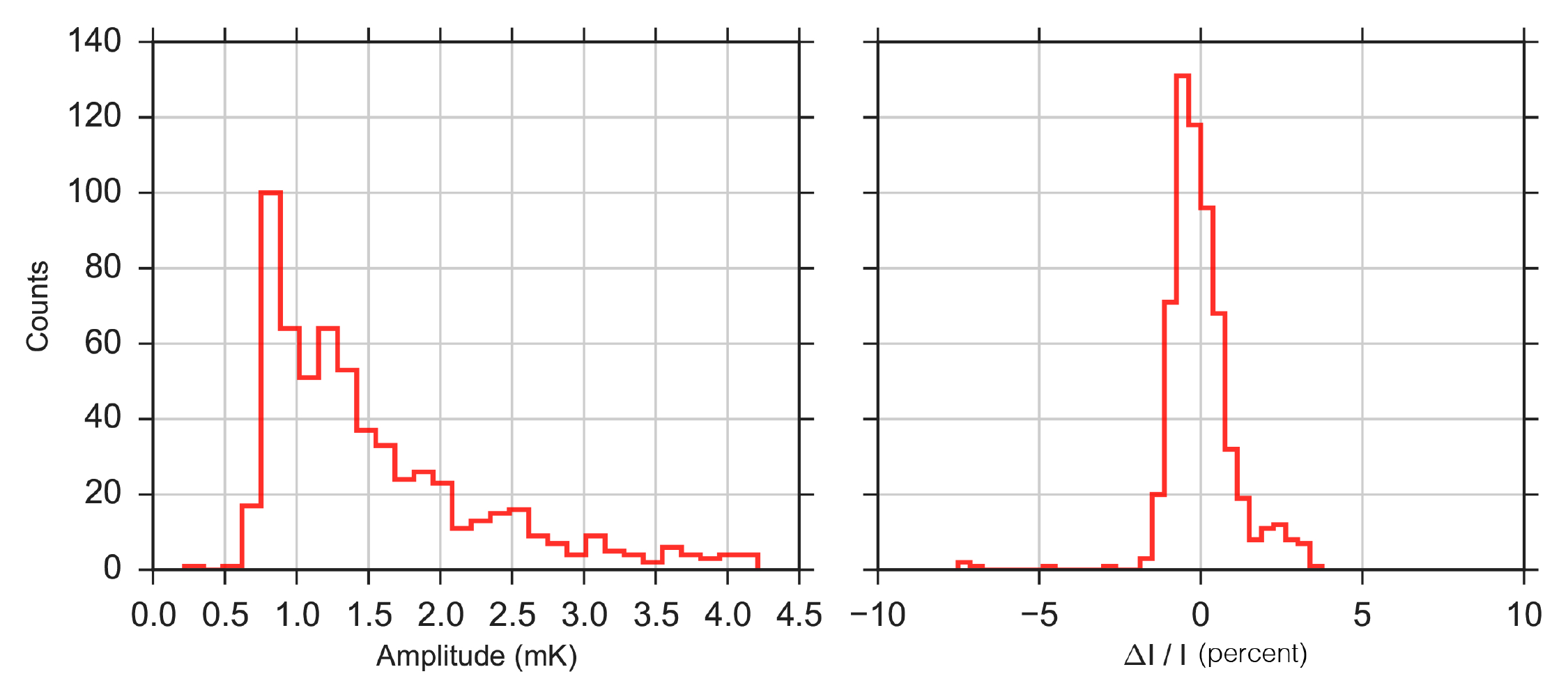}
\caption{\textit{Left:} Spectra of maximum amplitude for each pulse in the data set. \textit{Right:} Energy integrals as a percentage of the average value, normalised by the .8 pJ $^{241}$Am $\alpha$ emission line.}
\label{spectrabef}
\end{figure}

For all data, the maximum amplitude spectra of the dataset was compared with the energy spectra (represented as the total integral of each pulse~\cite{torres2012towards}) in Fig.~\ref{spectrabef}. We show that the pulse amplitudes have a wide distribution with a skew toward low amplitudes, which we interpret to be an effect of the position of the $\alpha$ particle impact on the absorber. Because of the relative size of the absorber with respect to the sensor, lower-amplitude events are statistically the most likely case. \\


As all of the $\alpha$ particle energy will be deposited into the absorber due to the stopping power range~\cite{berger1998stopping}, as well as the lack of thermal bath coupling on the absorber, all of the pulse energy must travel though the absorber before being read as a signal on the sensor. Thus, it is expected that the total energy deposited in each pulse will remain roughly constant, corresponding to the .8 pJ line. This is also noted in Fig.~\ref{spectrabef}, where the distribution of energy spectra shows a strong peak (normalised in the figure) with a FWHM of about 2\%. This finding corresponds to similar studies, which have shown that the energy deposition is represented by the total integral~\cite{torres2012towards} in the case where energy deposition is constant (as is the case with a single-energy $\alpha$ particle source).\\

We note a small irregularity in the energy spectra in Fig.~\ref{spectrabef} (right), just above $\Delta I / I = 0$ - we believe this feature may be due to a temperature instability late in the experiment, which would affect the working point of the bolometer, and therefore the total integrals the affected pulses.\\

Both of the above analyses are common to both data treatment algorithms used in the text - these results are obtained directly from preprocessed data, irrespective of later analytical techniques.\\

\subsection{Limitations of standard numerical analysis}
\label{sec:oldfits}
\begin{figure}[htbp]
\centering
\includegraphics[width=0.9\linewidth, keepaspectratio]{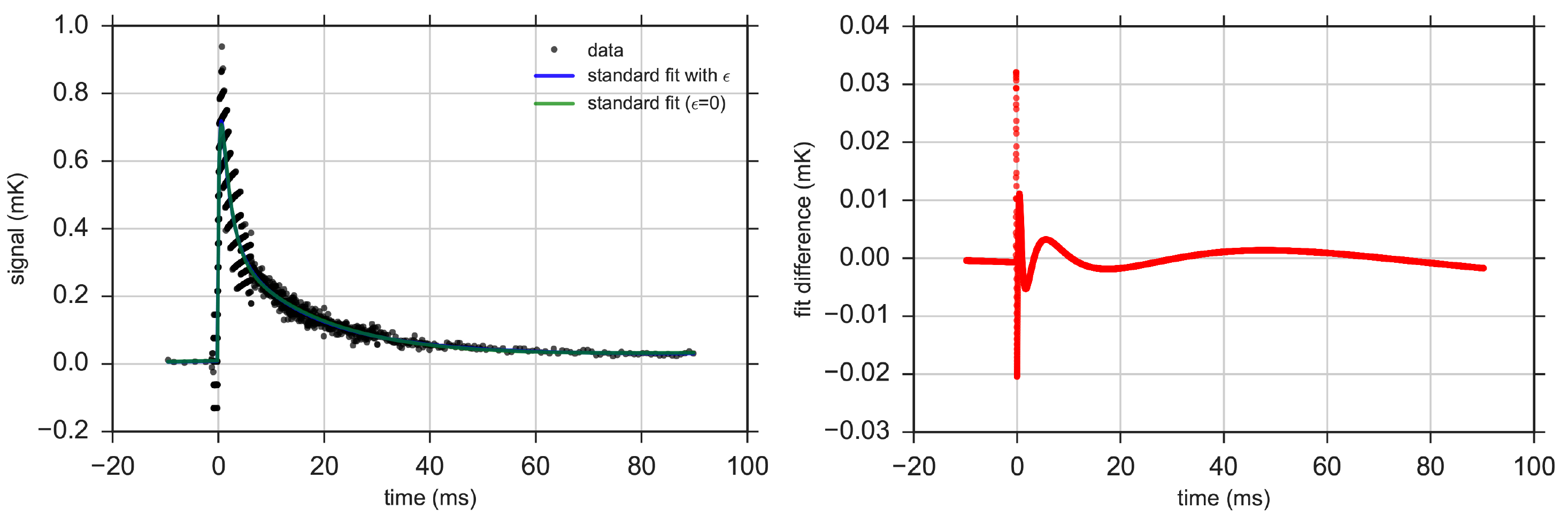}
\caption{\textit{Left:} One sample pulse (black) using the typical double-exponential fit (blue) and the same for $\epsilon = 0$ (green). \textit{Right:} The difference between the two left fits (with free $\epsilon$ and $\epsilon = 0$.)}
\label{oldfitresi}
\end{figure}

Previous analyses of pulse shapes in similar experiments~\cite{d2016cryogenic} typically employ the sum of two double exponentials, in the form of:

\begin{equation}
f_{glitch}(t, t > 0) = A_1\times\left( e^{-t/\tau_{2}}-e^{-t/\tau_{1}}\right)+A_2\times\left( e^{-t/\tau_{4}}-e^{-t/\tau_{3}}\right)
\label{fiteq}
\end{equation}

where A$_{1}$ and A$_{2}$ are the amplitudes of the athermal and thermal phonon components (respectively), $\tau_{1}$ and $\tau_{2}$ are the athermal phonon rise and decay times, and $\tau_{3}$ and $\tau_{4}$ are the thermal rise and decay times. For our usage, this equation was convolved with the electronic bandpass of the system, which set as a constant from an average calculated value of 0.027 ms. This fitting applied to one sample pulse, is shown in Fig~\ref{oldfitresi} (left).\\

\begin{figure}[htbp]
\centering
\includegraphics[width=0.5\linewidth, keepaspectratio]{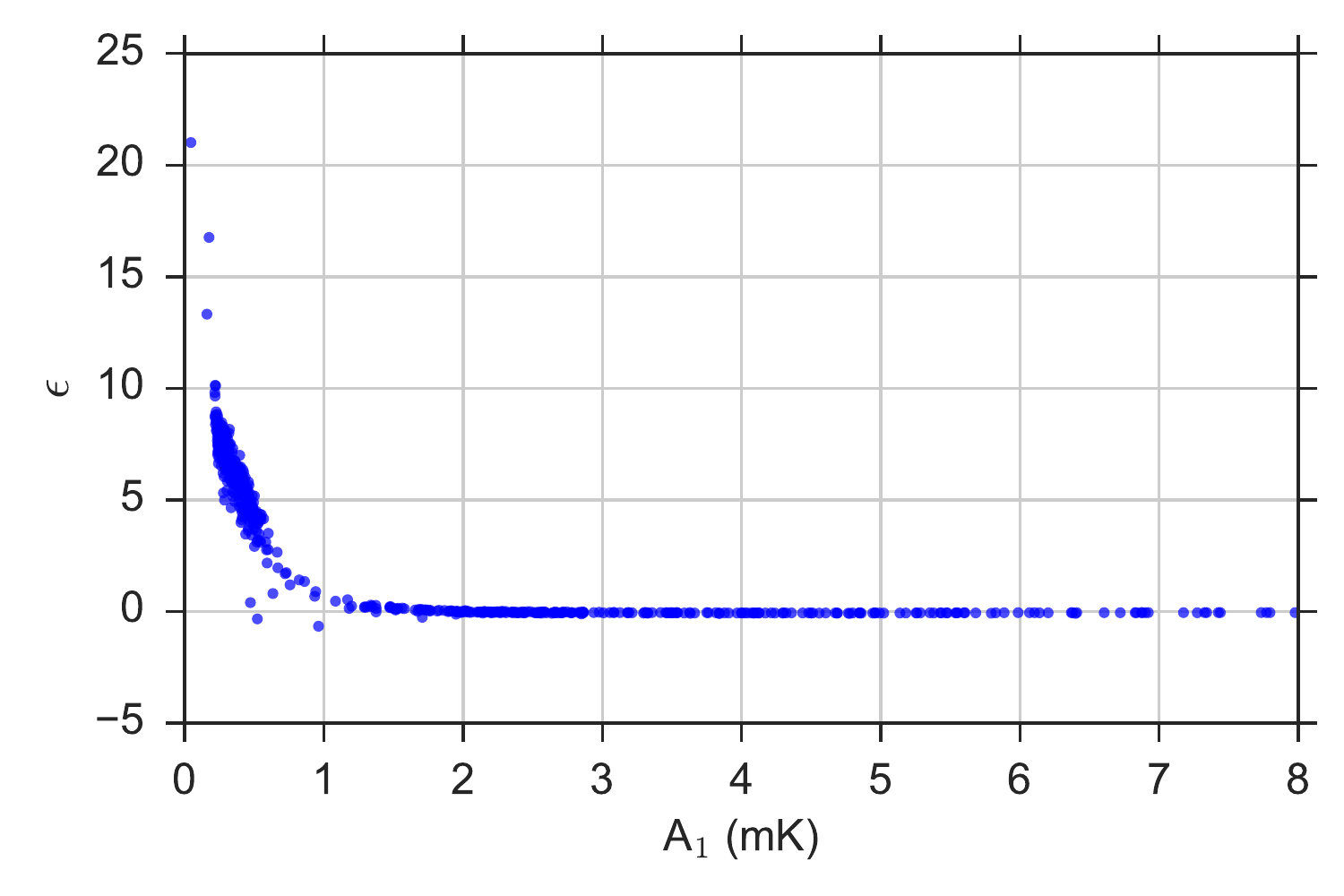}
\caption{The quadratic nonlinearity factor $\epsilon$ as a function of the ballistic amplitude A$_{1}$.}
\label{oldfitNL}
\end{figure}

The chi-squared goodness of the fits were improved (particularly over the smallest pulses) by the introduction of a quadratic nonlinearity factor, in the form of:

\begin{equation}
f_{glitch}' = f_{glitch} (1 + \epsilon f_{glitch})
\label{NLeq}
\end{equation}

which was well-correlated with amplitude; $\epsilon$ is very large (as high as 20\%) for low amplitude pulses, and small and negative for high amplitude pulses. We show the significant effect of epsilon on our sample fit in Fig~\ref{oldfitresi} (right), where we have the difference between the best fit with $\epsilon$ and the best fit where $\epsilon = 0$. We show $\epsilon$ as a function of amplitude in Fig.~\ref{oldfitNL}. We also note that this sample fit is one with a positive $\epsilon$, but only 4.65\% - the smallest fits can reach a much higher $\epsilon$, where there effects are even more pronounced.\\

Originally, $\epsilon$ was assumed to be related to the large, highly-localised energy distribution in the area of pulse impact, and the resulting nonlinearities in the detector. $\epsilon$ was one of the original motivations for investigation of these pulses via modelling - however, it was unable to be physically validated, proving that it was necessary to dig deeper into the physical origin of the pulse shape. In particular, we must explain the statistical attributes of the pulses using the physics of heat propagation, as well as the geometrical realities of the detector. \\

\begin{figure}[htbp]
\centering
\includegraphics[width=0.9\linewidth, keepaspectratio]{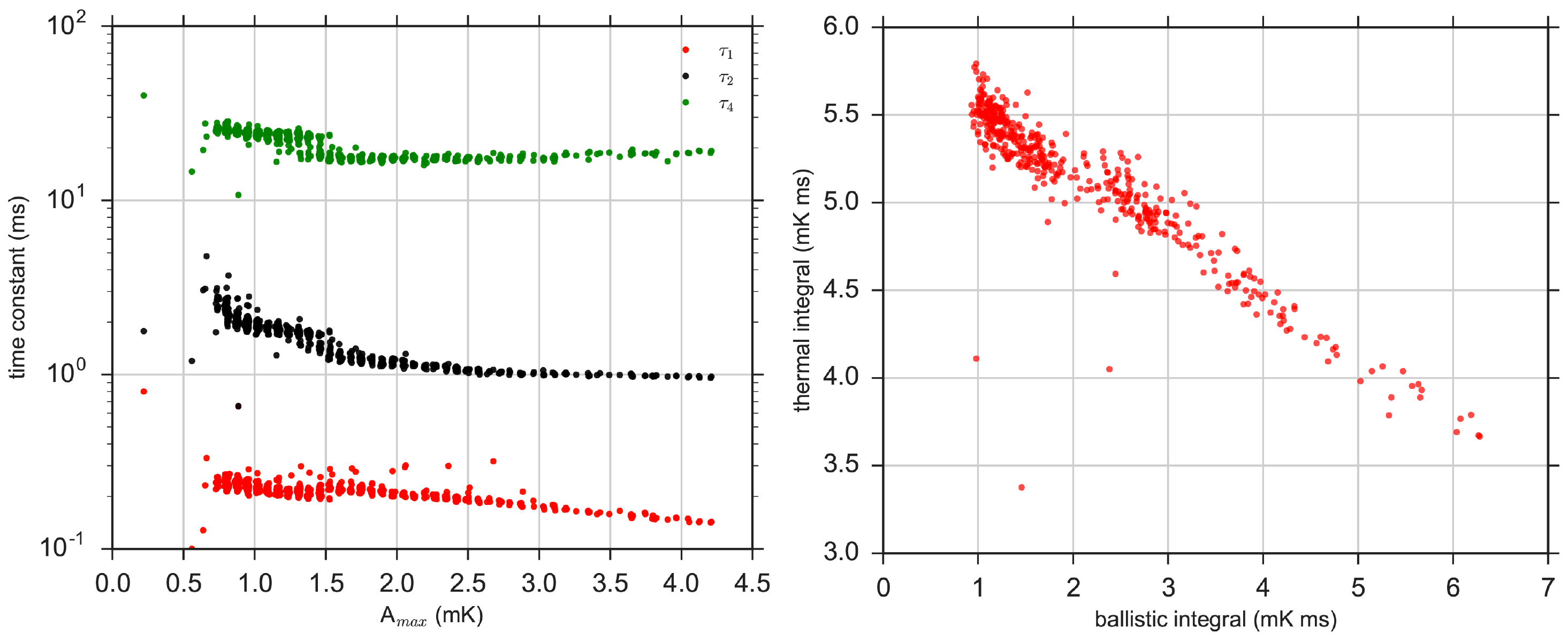}
\caption{\textit{Left:} Plots of the four time constants found by the double-exponential fit as a function of the maximum pulse amplitude A$_{max}$: $\tau_{1}$ (red) : Ballistic component rise time; $\tau_{2}$ (black) : Ballistic component decay constant, which is equal to the thermal rise time; $\tau_{4}$ (green): Thermal component decay time. \textit{Right:} Separated component integrals (ballistic vs. thermal) as a function of each other.}
\label{oldtauints}
\end{figure}

For the sake of comparison, we show the time constants and the ratios of the ballistic and thermal integrals for this dataset using this classical mathematical model in Fig.~\ref{oldtauints} (left). We note that the behaviour is as one would expect; the smallest pulses have the largest rise times ($\tau_{1}$). The ballistic and thermal integrals have a strongly linear relationship (seen in Fig.~\ref{oldtauints} - right), with the majority of the pulses being dominated by the thermal component due to the position effects mentioned previously.\\

The typical mathematical fit applied to pulse analysis in the literature provides a good fit to most pulses, as we can see in the residuals. The addition of $\epsilon$ improves the $\chi^{2}$ goodness of the fit, particularly for the smallest pulses, where $\epsilon$ aids in resolving the rise time of the pulse - this portion of the pulse is the most important for understanding the effect of ballistic phonons. The average full $\chi^{2}$ over the dataset without $\epsilon$ is 2209, and with $\epsilon$ is 2146 - an effect which dominates in the small-pulse regime, as we have shown in Fig.~\ref{oldfitNL}. However, this significant component does not appear to arise from any physical phenomena, which is necessary to quantify in order to truly understand the pulse shape. \\

\subsection{New analytical fitting algorithm} 
\label{sec:fitalgo}

With the thermal physics involved in the experiment in mind, we have built a new analytical model for the pulse shape. This new model has been developed to reproduce the real pulse shapes without the non-physical nonlinearity coefficient $\epsilon$, and to quantify heat propagation on the absorber disc with a new parameter extracted from the fit, for each pulse. We plan to use the same fitting model on simulated pulses to deepen our understanding of this bolometer. This section presents the basic ingredients of this fitting function.\\
 
When a 5.4 MeV $\alpha$ particle impacts the absorber disc, it loses all its kinetic energy, which is deposited through ionisation concentrated at the end of its trajectory in the absorber, after an average range of 14$\mu$m~\cite{berger1998stopping}. This energy is converted locally into ballistic phonons~\cite{yvon1996evidence} which radiate isotropically from the original impact area at high speed (the speed of sound in the media is typically a few $km/s$) before thermalising on the top or bottom disc surface (or on its periphery) at distances typically roughly the size of the thickness of the absorber ($40\ \mu m$). We surmise that the thermalisation of ballistic phonons is very rapid compared to all other time constants in the bolometer. This first process gives a starting lateral heat profile on the disc which depends on the $\alpha$ impact point. In the central region of the absorber disc, the shape of the profile is invariant; however, this is not the case for $\alpha$s falling near the absorber disc border. In any case, it presents as a peaked core centred on the $\alpha$ impact point with a geometrical $1/x^{2}$ tail.\\
 
The closer (or further) distance of the impact point to the disc center results in a faster (or slower) thermal propagation time to the sensor, where the signal is read (a position effect which has been seen in other experiments~\cite{vaillancourt2004large}). Whilst it is likely true that `faster' peaks are the result of ballistic phonons which arrive closer to the detector sensor, we assert that the ballistic phonons themselves thermalise at surface boundaries or in the absorber disc, and that all energy passing through the sensor takes the form of thermal phonons. In the case of a particle impact immediately above the sensor, more ballistic phonons thermalise at the Bi/Ge interface, creating a larger initial pulse. As a consequence of this, we believe it is necessary to challenge the assertion that the first exponential comes from `ballistic energy' and that the second one comes from `thermal energy'. With this in mind, we propose an alternative to the standard double-exponential form of glitch analysis. For brevity, we will continue to use the usual `ballistic' and `thermal' terminologies to describe pulse components, amplitudes, and attributes; however, in the case of this detector, we assume that all athermal phonons relax to thermal phonons before they reach the sensor. The following process consists of heat diffusing on the absorber disc before moving through the sensor towards the heatsink.\\

We expect a potentially large heat propagation time in the thin diamond absorber (particularly for far-away pulses) due to Casimir predictions for thermal conduction in thin layers~\cite{casimir1938note}. These characteristic times are higher than the thermal coupling time constant from the central part of the absorber to the sensor. We simplify the absorber thermal coupling to the bolometer by considering
heat diffusion in the absorber and heat flow to the sensor as two independent processes. From this, we get a simple pulse shape description based on the convolution of both processes. In the ongoing modelling of this detector, we use a combination of two methods: (1) a Monte-Carlo particle propagation model for isotropically radiating ballistic phonons, which thermalise and then diffuse as thermal phonons, and (2) a thermal block model for the impulse response function of the sensor and sapphire layers. Using the same reasoning for the pulse fitting, we can also use two components: (1) $f_{therm}$, the temperature at the centre of an isolated absorber, and (2) $f_{response}$, the bolometer response function. $f_{response}$ is defined as the sensor pulse caused by a temperature step on top of the sensor:\\

\begin{equation}
f_{response}(t)\equiv \frac{dH(t)}{dt} \otimes f_{response} =  \int_{0}^{t} \delta_{Dirac}(\tau) \times f_{response}(t - \tau)  d\tau
\label{conv}
\end{equation}

Applied to our general case, it gives the full model function:\\

\begin{equation}
T_{sensor}(t)= \frac{df_{therm}}{dt} \otimes f_{response} =  \int_{0}^{t} \frac{df_{therm}(\tau)}{dt} \times f_{response}(t - \tau)  d\tau
\label{conv}
\end{equation}

To be able to quickly compute this function, we make further approximations in the evolution of the temperature profile. In the temperature domain used, the diamond heat capacity and thermal conduction are both dominated by their $T^3$ dependence, which cancels out of the heat propagation differential equation. Therefore, temperature evolves like in the constant $C$ and $G$ case.\\

If we approximate the initial heat profile with a 2-dimensional Gaussian, we can rely on the properties of standard heat propagation. Noting $\Delta T(\rho,t)$, the differential temperature profile centred on the $\alpha$ impact position, we have: 

\begin{equation}
\Delta T(\rho,t) = \frac{A}{t+t_{dep}}e^{-a\rho^2/(t+t_{dep})}
\label{eqdiff}
\end{equation}

where the lateral size increases with the square root of time, and $t_{dep}$ represents spread of the initial heat profile. $a\rho^2$ has the dimension of time, and we denote it $\tau_3$. It is expected to depend on the distance $\rho$ between the $\alpha$ impact and the absorber disc centre, increasing like $\rho^{2}$ in this simplified picture.\\

This approximation is suited for an infinite plate. A simple way to take the small disc dimension into account is to keep only the exponential dependence, so $\Delta T$ reaches a nonzero asymptotic value after a long period of time, as expected for an isolated absorber. Finally, on top of the sensor, we have:

\begin{equation}
f_{therm} = H(t)\ A_{therm} e^{-\tau_3/(t+t_{dep})} {\rm ,\ with\ }H(t)=0  {\rm\  for\ } t<0\ {\rm\  and\ }H(t)=1 
{\rm\  for\ } t\ge 0
\label{eqdiff2}
\end{equation}

For $f_{response}$, we take a standard 2-component exponential response function which was produced by the thermal block model of the bolometer. The complete function contains 11 parameters, outlined in Table~\ref{params}.\\

\begin{table}[]
\centering
\caption{Table of new function parameters and their descriptions.}
\label{params}
\begin{tabular}{|l|l|}
\hline
{\textbf{parameter name}} & {\textbf{parameter description}} \\ \hline
$\tau_1$ &  rise time (response function) \\ \hline
$\tau_2$ & Thermal pulse decay time (response function) \\ \hline
$\tau_3$ & Thermal energy propagation time \\ \hline
$\tau_4$ & Athermal pulse decay time (response function) \\ \hline
t$_{deposition}$ & Initial thermal profile spread \\ \hline
A$_{therm}$ & Amplitude of thermal profile \\ \hline
A$_{1}$ & Amplitude of thermal pulse \\ \hline
A$_{2}$ & Amplitude of ballistic pulse \\ \hline
offset & Oscilloscope / data offset  \\ \hline
slope & Artificial slope induced by filtering technique \\ \hline
t$_{trigger}$ & Pulse trigger time (time offset) \\ \hline
\end{tabular}
\end{table}

These parameters represent the initial thermal profile of amplitude $A_{therm}$ with a width proportional to $t_{dep}$, spreading over the disc with a characteristic time constant $\tau_{3}$, and a standard two-component exponential response function (where the amplitudes and the time constants have their usual meanings). \\

To decrease computation time, we can set $\tau_{1}$ to an average calculated value of 0.08 ms in order to be short enough to compensate for the fastest high-amplitude pulses, allowing $\tau_{3}$ to compensate for lower rise times. We also set A$_{therm}$ to 1 to normalise the amplitude of the thermal profile in the convolution. Using this technique, we can compare the output of this function with fits from the typical numerical method~\cite{d2016cryogenic} we have already shown in Sec.~\ref{sec:oldfits}, e.g. by splitting the total pulse into separate `ballistic' and `thermal' components, comparing the relationship of their integrals as an approximation for initial $\alpha$ impact position, etc.\\

\section{Output and Analysis of new fitting algorithm} 
\label{sec:analysis}

In this section, we show the output of fits with the new pulse shape, comparing this with the results of the commonly-used methods (shown in Sec.~\ref{sec:oldfits}).

\subsection{Goodness of fits} 
\begin{figure}[htbp]
\centering
\includegraphics[width=0.95\linewidth, keepaspectratio]{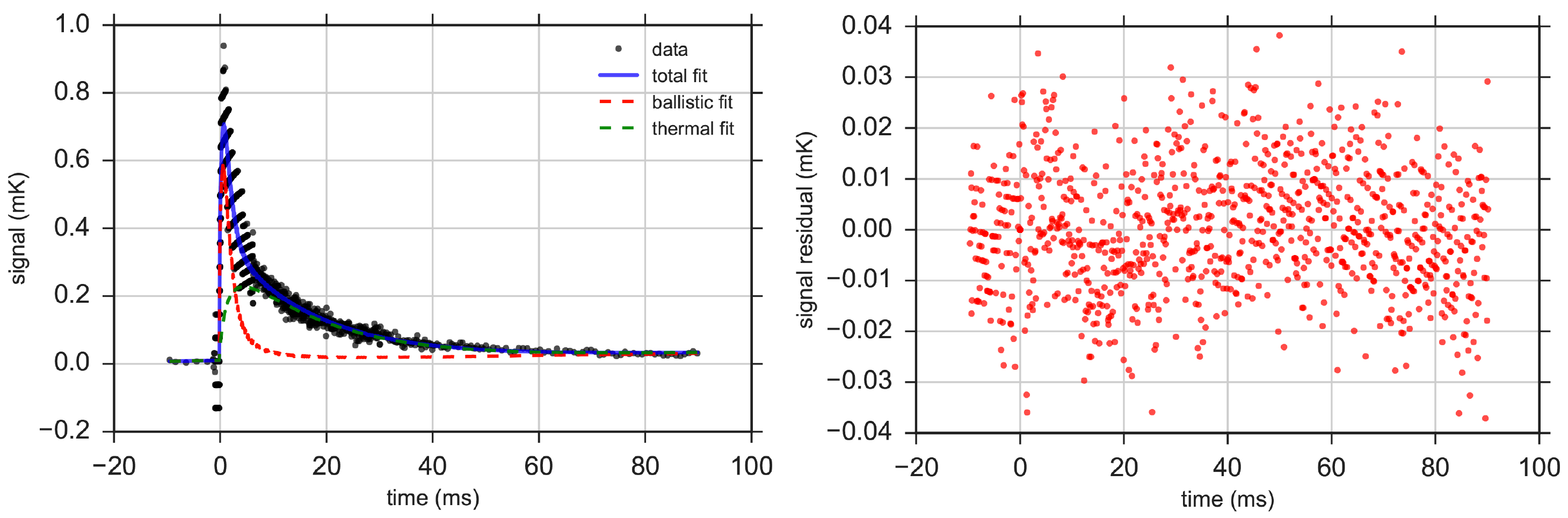}
\caption{\textit{Left}: A sample pulse (data in black), total pulse fit (blue), ballistic pulse fitting (red) and thermal pulse fitting (green). \textit{Right}: The evenly-sampled residuals of the total fit of the same unpacked pulse, with the plotted residuals sampled every 100$\mu$s.}
\label{fitresi}
\end{figure}

Employing the new pulse shape described in Sec~\ref{sec:fitalgo}, we are able to fit to the total pulse, as well as the `fast' and `slow' components, as we have done in earlier sections. We find an average $\chi^{2}$ per degree of freedom of 1.13. We note that the error in this fitting appears to be dominated by measurement noise sources, particularly a high amount of digitisation noise in the pulses (of the order of 0.1 V), which has been attributed to the oscilloscope and readout chain. In analysing the residuals of the fitted pulses, we find that the residuals are evenly distributed about 0 mK, and we are therefore primarily limited by noise rather than parametric uncertainty. Comparing with the fits of the typical mathematical model outlined in Sec.~\ref{sec:oldfits}, we note that the average $\chi^{2}$ appears to be roughly equivalent to that of the typical fitting method (with an average $\chi^{2}$ per degree of freedom of 1.012). In some cases, the artificial slope induced by the filtering technique is prominent for both fitting methods, but has been accounted for in both fitting processes, and does not appear in the residuals. These features are demonstrated in Fig.~\ref{fitresi}.

\subsection{Statistical data features} 
In further processing the pulses as shown in Fig.~\ref{fitresi}, we can compare the fit amplitudes as well as the time constants of each pulse (with the exception of the response function rise time, $\tau_{1}$, which is fixed). Finally, we can also compare the integral of the measured pulses with those of the split fast and slow components of the response functions, to understand the distribution of energy in each pulse, following the comparative treatment using the typical fitting method in Sec~\ref{sec:oldfits}.\\

\begin{figure}[htbp]
\centering
\includegraphics[width=0.8\linewidth, keepaspectratio]{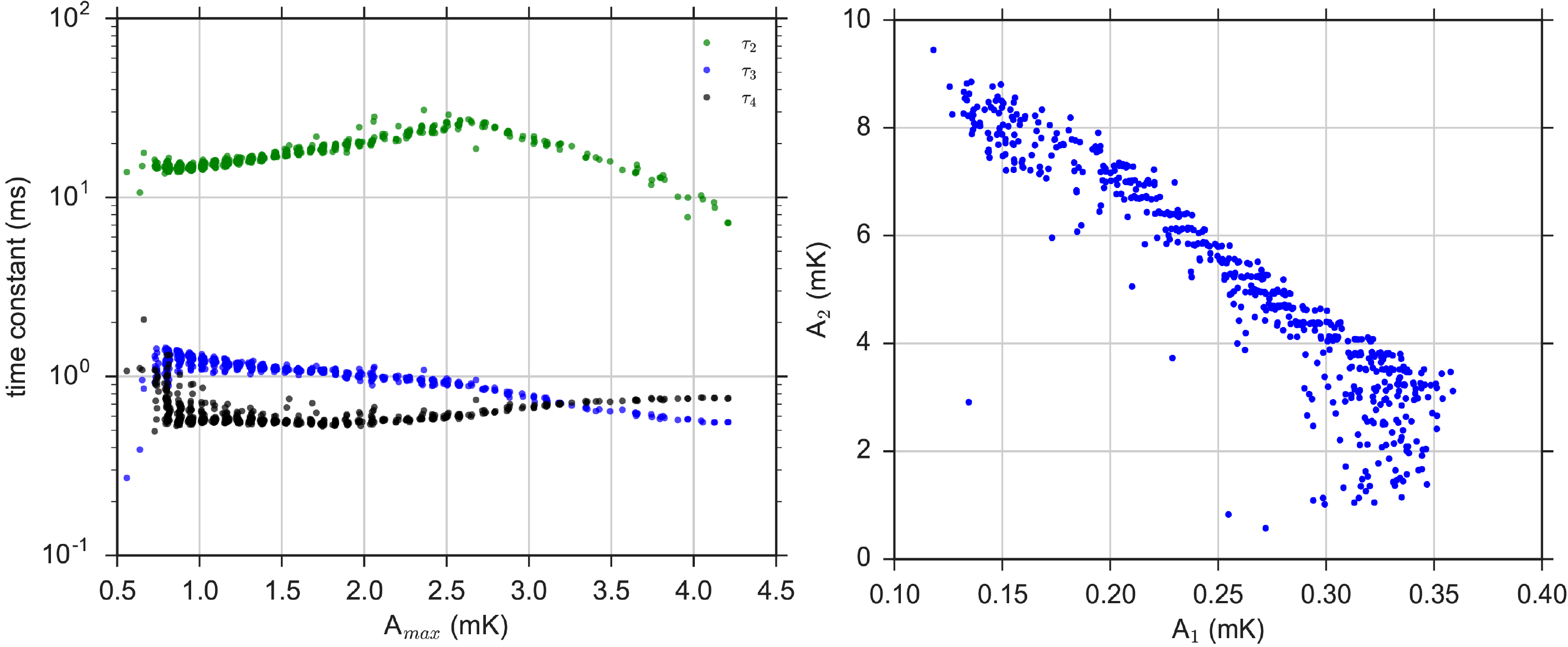}
\caption{\textit{Left:} Plots of the four time constants as a function of the maximum pulse amplitude A$_{max}$: $\tau_{2}$ (green) : Slow response function time constant; $\tau_{3}$ (blue): Thermal profile time constant; $\tau_{4}$ (black): Fast response function decay time. \textit{Right:} Response function amplitudes A$_{2}$ (ballistic) and A$_{1}$ (thermal) as a function of each other. }
\label{taus}
\end{figure}

In Fig.~\ref{taus} (left), we show the relationships between the maximum amplitude of each pulse and the time constants produced by the fitting. The shortest time constant, $\tau_{1}$, is fixed to 0.079 ms to account for the fastest (highest-amplitude) pulses. Pulses which rise slower than this are compensated for in the convolution with the thermal profile. The thermal propagation time constant, $\tau_{3}$, appears to be strongly correlated with the thermal response function decay constant $\tau_{4}$ - they rise and decay inversely with one another, indicating regions in parameter space where the effects of thermal propagation become more or less prolific than the response function decay effects. We note that the maximum amplitudes are affected by digitisation noise coming from the oscilloscope and the low V$_{bias}$ of the experiment. \\

\begin{figure}[htbp]
\centering
\includegraphics[width=0.9\linewidth, keepaspectratio]{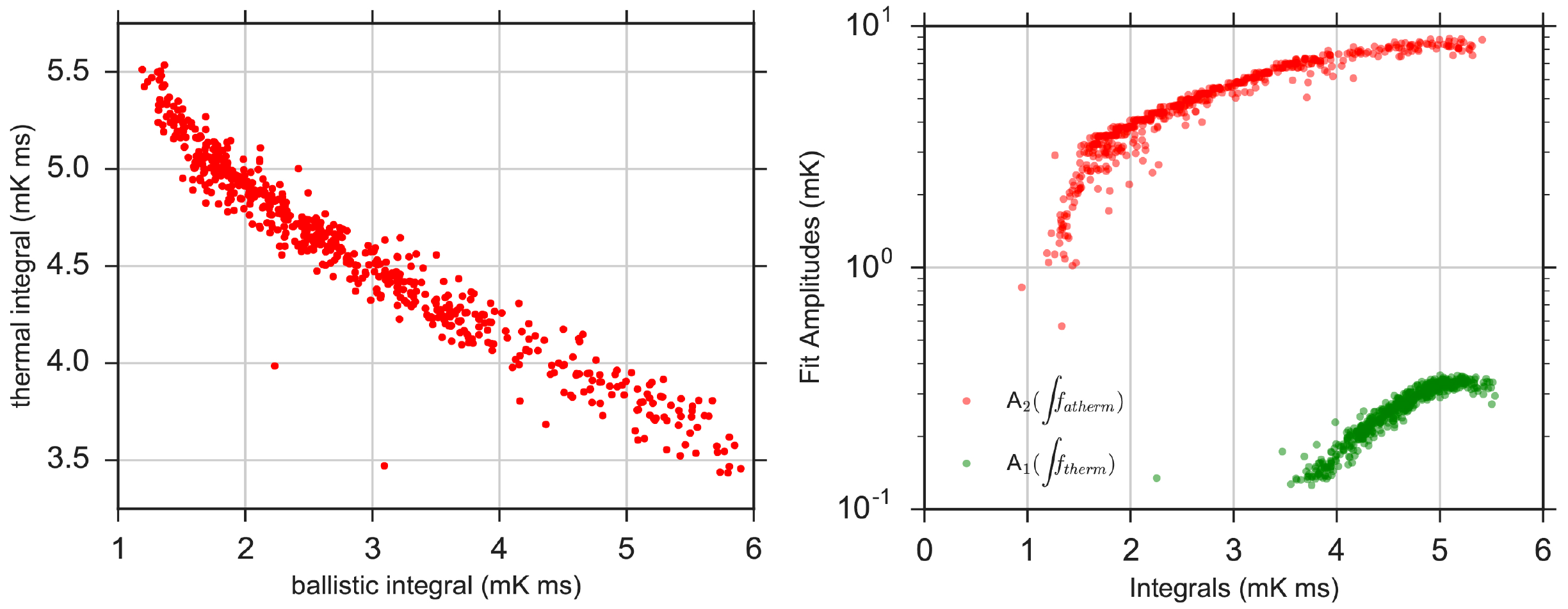}
\caption{\textit{Left}: The relationship between athermal (fast) and thermal (slow) integrals. \textit{Right:} Response function amplitudes as a function of their respective (thermal or ballistic) integral.}
\label{ampsints}
\end{figure}

We also see in Fig.~\ref{taus} (right) the relationships between the two fit amplitudes A$_{1}$ (the slow component arising from propagation of thermal energy across the absorber) and A$_{2}$(the fast component arising from thermalised athermal phonons). For the most part, the relationship between these two amplitudes appears to be linear, showing the expected sharing of energy between the two components; however, that relationship appears to break down at high A$_{1}$, which are the impacts most likely to be on the periphery of the absorber. In these pulses, A$_{2}$ tends toward 0 more quickly than expected. We note that the fit amplitudes are not directly linked to the maximum pulse amplitude A$_{max}$ - it also depends on $\tau_{3}$, and divergences may be an effect of the convolution.\\


Fig.~\ref{ampsints} (left) demonstrates the relationship between the thermal and ballistic integrals. We find a correlation which is mostly linear, with exception for the highly-thermal pulses which we have mentioned previously. This relationship is very similar to that which we find using the classical mathematical method outlined in Sec.~\ref{sec:oldfits} -- the linearity of these two components serves as a validation measure for the new fits, demonstrating that the overall distribution of component ratios is much the same, owing largely to position effects, with only a small divergence for the furthest $\alpha$ impacts. Fig.~\ref{ampsints} (right) shows the relationship between the thermal (green) and athermal (red) components of the fit amplitudes with the thermal and athermal integrals (respectively), showing a relationship which is mostly linear, but is affected by the same divergences at high A$_{1}$ and low A$_{2}$.\\

\begin{figure}[htbp]
\centering
\includegraphics[width=0.5\linewidth, keepaspectratio]{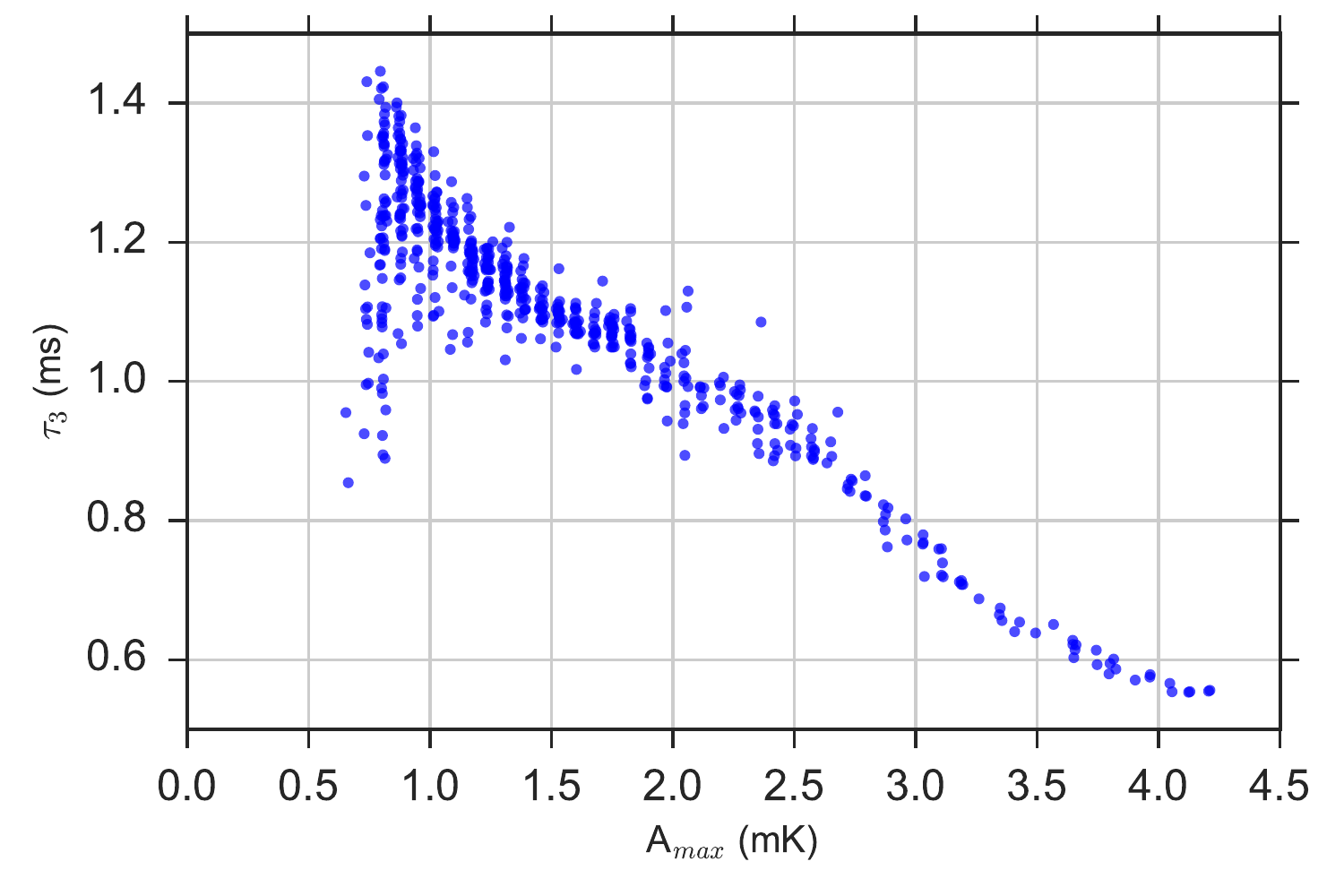}
\caption{Linearly scaled $\tau_{3}$ as a function of the maximum amplitude.}
\label{tau3}
\end{figure}

Finally, we show the relationship between the maximum pulse amplitude and $\tau_{3}$, in linear scale for clarity, in Fig.~\ref{tau3}. We expect a potentially large propagation time in the thin diamond absorber (particularly for far-away pulses) due to Casimir predictions for thermal conduction in thin layers~\cite{casimir1938note}. We find that this relationship is mostly linear until low A$_{max}$, where the dispersion increases inversely with amplitude. \\


\subsection{Discussion}
\label{sec:discussion}
Using the new fitting method as a means of testing the current physical understanding of energy propagation in this bolometer, we find some validation of our physical interpretation. In particular, we find that the convolution of a thermal input with the bolometer response function appears to be a valid way to describe $\alpha$ particle pulses in this detector. Pulses arising from alternative particle energy deposition (e.g. minimally-ionising protons) is planned in order to validate these theories, and will be the subject of future study.\\

The results of the model appear to be equivalent to the previous fitting methods in terms of agreement with the data, but represent an improvement due to the lack of necessity for the nonlinearity factor $\epsilon$, which has not been physically quantified. The new fitting method appears to produce similar overall behaviour to that of the sum of two double exponentials, but is based on the principles of thermal physics, rather than on a mathematical expression alone. The ratio of the total energies in the fast and slow pulse components remains similarly position-dependent, lending credibility to the idea that the pulse shape is mostly a result of the position of the $\alpha$ particle's impact on the absorber, which creates differences in the way energy propagates before being measured as a signal in the sensor. \\

When the $\alpha$ particle impacts the absorber, the distance of the barycentre of that impact is the main driver behind how much of its energy is distributed into the fast and slow pulse components. In the numerous cases where the $\alpha$ impact is closer to the absorber border than to the sensor, many ballistic phonons will likely thermalise at the border of the sensor, and the thermal phonons which would normally have a Gaussian distribution in the absorber will be `stacked' at the border. Those thermal phonons will eventually travel to the sensor, but less quickly than the energy distributed closer to the sensor. When a large amount of energy is initially stored at the border, that energy can appear as a second `wave' which arrives at the sensor some time after the initial energy spike in the `fast' component. We see this effect in $\tau_{3}$ and $\tau_{4}$ - their behaviours are complimentary at very low and very high initial amplitudes. The effect of the thermal propagation time $\tau_{3}$ is particularly prevalent when compensating for pulses with a very small ballistic component, where a large amount of energy is temporarily stored at the periphery of the absorber. However, $\tau_{3}$ also displays a significant improvement on the high (nearby to sensor) pulses; a nonzero $\tau_{3}$ allows for better resolution of the initial pulse rise and decay times, compared with a zeroed $\tau_{3}$, which is mathematically analogous to the classical mathematical model used previously.\\

The fitting algorithm is also computationally intensive, and requires highly-sampled data to be resampled to save computational time. The new algorithm does not show an increase in the $\chi^{2}$ goodness, but is roughly equivalent, and with a greater physical significance.\\

\section{Conclusions}
We have presented a new pulse shape description for $\alpha$ particles in a composite NTD Ge bolometer with a large, disc-shaped absorber, using a dataset of pulses at 100 mK as a proof-of-concept, comparing with the typical mathematical analysis methods. The new algorithm is based on the physical understanding of this detector and the means of energy propagation within it, and contains a convolution of a profile of thermalised ballistic phonons (based on a gaussian spread of thermal energy in the x-y plane) and a classic bolometer response function (the sum of two double exponentials). The motivation for the new algorithm is based on the limitations of previous mathematically-motivated fitting methods\cite{d2016cryogenic}, which were in good agreement with this dataset, but which necessitated the use of a quadratic nonlinearity factor resolve the fastest pulses. This nonlinearity factor, $\epsilon$, was not able to be physically validated in modelling or in experiment. We have therefore expanded upon the previous fitting algorithm by employing a convolution with a temperature profile, which accounts for the movement of thermal energy from an initial distribution on the absorber region of impact.\\

We have tested both algorithms with a data set of $\alpha$ particle impacts at 100 mK. The data has been finely sampled and must be processed before applying the algorithm, to decrease computational time and reduce significant noise (an effect of a low bias voltage). The results are similar to that which were achieved using the numerical function, showing that the ratio of athermal and thermal energies and amplitudes is largely unchanged. The parameters of the new algorithm are more well-constrained than the previous methods, although limitations are shown in the regime of low-amplitude pulses ([maybe i can take this out when i fix the fit]). Further improvements must be made on pulses in this regime for further analysis. \\


\acknowledgements     
 
The author wishes to acknowledge the french national space agency Centre National d'Etudes Spatiales (CNES) for funding their work. 


\bibliography{bib}   

\begin{thebibliography}{10}

\bibitem{lamarre2010planck}
J.-M. Lamarre, J.-L. Puget, P.~A. Ade, F.~Bouchet, G.~Guyot, A.~Lange,
  F.~Pajot, A.~Arondel, K.~Benabed, J.-L. Beney, {\em et~al.}, ``Planck
  pre-launch status: The hfi instrument, from specification to actual
  performance,'' {\em Astronomy \& Astrophysics}~{\bf 520}, p.~A9, 2010.

\bibitem{catalano2014characterization}
A.~Catalano, P.~Ade, Y.~Atik, A.~Benoit, E.~Br{\'e}ele, J.~Bock, P.~Camus,
  M.~Charra, B.~Crill, N.~Coron, {\em et~al.}, ``Characterization and physical
  explanation of energetic particles on planck hfi instrument,'' {\em Journal
  of Low Temperature Physics}~{\bf 176}(5-6), pp.~773--786, 2014.

\bibitem{catalano2014impact}
A.~Catalano, P.~Ade, Y.~Atik, A.~Benoit, E.~Br{\'e}ele, J.~Bock, P.~Camus,
  M.~Chabot, M.~Charra, B.~Crill, {\em et~al.}, ``Impact of particles on the
  planck hfi detectors: Ground-based measurements and physical
  interpretation,'' {\em Astronomy \& Astrophysics}~{\bf 569}, p.~A88, 2014.

\bibitem{benoit2000calibration}
A.~Benoit, F.~Zagury, N.~Coron, M.~De~Petris, F.-X. D{\'e}sert, M.~Giard, J.-P.
  Bernard, J.-P. Crussaire, G.~Dambier, P.~De~Bernardis, {\em et~al.},
  ``Calibration and first light of the diabolo photometer at the millimetre and
  infrared testa grigia observatory,'' {\em Astronomy and Astrophysics
  Supplement Series}~{\bf 141}(3), pp.~523--532, 2000.

\bibitem{Janssen2018}
R.~M.~J. {Janssen}, S.~L. {Stever}, G.~{Rouille}, V.~{Sauvage}, M.~{Bouzit},
  and B.~{Maffei}, ``{Commissioning of a common-user test facility to evaluate
  the effects of high-energy particles on next-generation cryogenic
  detectors},'' in {\em Space Telescopes and Instrumentation 2016: Ultraviolet
  to Gamma Ray},  {\em Proceedings of SPIE Space Telescopes and Instrumentation
  2018} {\bf these proceedings}, 2018.

\bibitem{torres2012towards}
L.~Torres, N.~Coron, P.~de~Marcillac, M.~Martinez, and T.~Redon, ``Towards an
  absolute determination of the particle energy thermalized in bolometers,''
  {\em Journal of Low Temperature Physics}~{\bf 167}(5-6), pp.~961--966, 2012.

\bibitem{berger1998stopping}
M.~J. Berger, J.~Coursey, M.~Zucker, J.~Chang, {\em et~al.}, {\em
  Stopping-power and range tables for electrons, protons, and helium ions},
  NIST Physics Laboratory Gaithersburg, MD, 1998.

\bibitem{d2016cryogenic}
M.~D'Andrea, A.~Argan, S.~Lotti, C.~Macculi, L.~Piro, M.~Biasotti, D.~Corsini,
  F.~Gatti, and G.~Torrioli, ``The cryogenic anti-coincidence detector for
  athena x-ifu: pulse analysis of the ac-s7 single pixel prototype,'' in {\em
  SPIE Astronomical Telescopes+ Instrumentation},  pp.~99055G--99055G,
  International Society for Optics and Photonics, 2016.

\bibitem{yvon1996evidence}
D.~Yvon, L.~Berg{\'e}, L.~Dumoulin, P.~De~Marcillac, S.~Marnieros, P.~Pari, and
  G.~Chardin, ``Evidence for signal enhancement due to ballistic phonon
  conversion in nbsi thin films bolometers,'' {\em Nuclear Instruments and
  Methods in Physics Research Section A: Accelerators, Spectrometers, Detectors
  and Associated Equipment}~{\bf 370}(1), pp.~200--202, 1996.

\bibitem{vaillancourt2004large}
J.~Vaillancourt, C.~Allen, R.~Brekosky, A.~Dosaj, M.~Galeazzi, R.~Kelley,
  D.~Liu, D.~McCammon, F.~Porter, L.~Rocks, {\em et~al.}, ``Large area bismuth
  absorbers for x-ray microcalorimeters,'' {\em Nuclear Instruments and Methods
  in Physics Research Section A: Accelerators, Spectrometers, Detectors and
  Associated Equipment}~{\bf 520}(1-3), pp.~212--215, 2004.

\bibitem{casimir1938note}
H.~Casimir, ``Note on the conduction of heat in crystals,'' {\em Physica}~{\bf
  5}(6), pp.~495--500, 1938.

\end{thebibliography}
\bibliographystyle{spiebib}   

\end{document}